\documentstyle[12pt]{article}

\newlength{\dinwidth}
\newlength{\dinmargin}
\def\lapproxeq{\lower .7ex\hbox{$\;\stackrel{\textstyle
<}{\sim}\;$}}
\def\gapproxeq{\lower .7ex\hbox{$\;\stackrel{\textstyle
>}{\sim}\;$}}

\begin{document}
\titlepage
\begin{flushright}
DTP/96/22  \\
RAL-TR-96-019 \\
March 1996 \\
\end{flushright}

\begin{center}
\vspace*{2cm}
{\Large \bf Can partons describe the CDF jet data?} \\
\vspace*{1cm}
E.\ W.\ N.\ Glover$^a$, A.\ D.\ Martin$^a$, R.\ G.\ Roberts$^b$
and W.\ J.\ Stirling$^{a,c}$ \\

\vspace*{0.5cm}
$^a \; $ {\it Department of Physics, University of Durham,
Durham, DH1 3LE }\\

$^b \; $ {\it Rutherford Appleton Laboratory, Chilton,
Didcot, Oxon, OX11 0QX}\\

$^c \; $ {\it Department of Mathematical Sciences, University of Durham,
Durham, DH1 3LE }
\end{center}

\vspace*{4cm}
\begin{abstract}
The recent CDF single jet inclusive measurements at Fermilab are
incorporated in a global next-to-leading order parton analysis of
the available deep inelastic and related data.  We find that it
is impossible to achieve a simultaneous QCD description of both
the CDF jet distribution for transverse energies $E_T > 200$~GeV
and the deep inelastic structure function data for $x > 0.3$. 
However, the CDF data for $E_T < 200$~GeV and the deep inelastic
data are adequately described provided that the QCD coupling
$\alpha_S(M_Z^2)$ is increased from its preferred deep inelastic
value to $\alpha_S (M_Z^2) \simeq 0.116 - 0.120$.
\end{abstract}

\newpage
The measurement of the differential cross section for inclusive
central jet production at the Fermilab $p\overline{p}$ collider
has recently been reported by the CDF collaboration \cite{cdf}
for jet transverse energies, $E_T$, in the range 15 to 440~GeV. 
At the higher $E_T$ values these measurements probe the
substructure of the proton in a previously unexplored kinematic
region, equivalent to 4-momentum transfer squared $Q^2 \sim
10^5$~GeV$^2$ which corresponds to distance scales of $O (10^{-
17}$~m) or less.  Intriguingly, the experimental results for $E_T
> 200$~GeV show evidence of a possible deviation above the
behaviour predicted by next-to-leading order (NLO) QCD based on
the current sets of parton distributions --- distributions which
are obtained from global analyses \cite{mrsa,cteq,mrsag} of a
wide range of deep inelastic and related data.  Clearly before
explanations based on New Physics \cite{np} can be taken
seriously, it is
crucial to see if the parton distributions can be adjusted to
accommodate the jet measurements whilst retaining a satisfactory
description of the other data.  Here we address this vital
question.

We begin by comparing the CDF jet measurements with the (NLO) prediction
obtained from the MRS(A$^\prime$) set of partons \cite{mrsag}
which, at present, seem best able to describe all the other data.
The calculation uses the next-to-leading-order parton level
Monte Carlo JETRAD \cite{jetrad} and the cuts
and jet algorithm applied directly to the partons
are modelled as closely as possible to the experimental
set-up. The jets are defined according to the 
Snowmass algorithm with a jet cone size $R=0.7$ \cite{Snowmass}, and are  
required to lie in the
pseudo-rapidity range between $0.1 < |\eta| <  0.7$.
The factorization
and renormalization scales are chosen to be $\mu_F = \mu_R = E_T/2$.
The fractional difference between data and theory
is given by comparing the data points
with the horizontal line in Fig.~1.\footnote{For clarity 
we have not made the small overall renormalization of the A$'$ prediction
which would improve the average absolute value, but not the
description of the {\it shape} of the jet distribution.}
  Over the observed
range of $E_T$ the experimental cross section falls by more than
8 orders of magnitude and the sum of the {\it correlated}
systematic uncertainties grows from $\pm$ 18\% at $E_T = 50$~GeV
to $\pm$ 28\% at $E_T = 300$~GeV.  However, the statistical
precision of the data is such that the difference between the
observed and predicted shapes of the $E_T$ distribution must be
studied seriously.  Although we have selected MRS(A$^\prime$), it
is important to note that the comparison is very similar for the
other parton sets obtained from global analyses which include both
fixed-target and 
HERA deep inelastic data, see Fig.~1 of Ref.~\cite{cdf}. 
Disregarding a small change in normalization, the nature of the
discrepancy between the MRS(A$^\prime$)-based prediction and the
data differs according to whether the jet $E_T$ is below or
above about 200~GeV.  Interestingly it is around this value that
jet production changes from being dominated by $qg$-initiated to
(valence quark) $q\overline{q}$-initiated QCD subprocesses, see
Fig.~2.

Before we present a global parton analysis which incorporates
both the deep inelastic and jet data we discuss some general trends
that can already be established, bearing in mind that a centrally
produced jet of transverse energy $E_T$ samples partons at
\begin{equation}
x \; \sim \; 2 E_T/\sqrt{s}, \qquad Q^2 \; \sim \; E_T^2
\label{eq:a1}
\end{equation}
where $\sqrt{s} = 1.8$~TeV at the Tevatron collider.  First we note that the
discrepancy in the shape of the jet distribution for $E_T < 200$~GeV
 could be removed either by significantly increasing the QCD
coupling from the MRS(A$^\prime$) value of $\alpha_S (M_Z^2) =
0.113$ (see Refs.~\cite{ewng,mrsv}) or by increasing the gluon
in the $x \sim 0.1$ region (see, for example, the
MRS(D$_0^\prime$) \cite{mrsd} \lq\lq base" line in Fig.~1 of 
Ref.~\cite{cdf}).  Both options lead to disagreement with data
elsewhere.  The effect  of increasing $\alpha_S$ is
to increase the rate of increase in the partons at small $x$ and
to steepen the low $E_T$ jet cross section as required,  but it also
leads to scaling violations of deep inelastic data which are too
rapid.  The second option is contrary to HERA data which require
the gluon to be larger (and steeper) than D$_0^\prime$ for $x
\leq 10^{-2}$ and hence, by momentum conservation, smaller (and
steeper) for $x \sim 0.1$, see Fig.~15 of Ref.~\cite{mrsa}.

Turning now to the discrepancy for $E_T > 200$~GeV, we note that
in this region the jet distribution depends on the partons
(mainly the valence quarks, but also the gluon) with $x \sim
0.35-0.5$
and $Q^2 \sim 10^4-10^5$~GeV$^2$.  The discrepancy has
been accommodated in a recent analysis \cite{jh} by modifying the
gluon distribution in this $x$ region.  Traditionally the gluon
is constrained in the region $0.35 \lapproxeq x \lapproxeq 0.55$
by the WA70 prompt photon data \cite{wa70}.  In Ref.~\cite{jh} it is argued
that the modified gluon can be made consistent with the WA70 data
by taking into account scale uncertainties and $k_T$ broadening. 
However, increasing the gluon in this $x$ range implies a decrease 
in the region
$x \sim 0.01-0.3$ (see Fig.~2 of \cite{jh}) so the
resulting partons do not describe the shape of the more precise
jet data with $E_T < 200$~GeV, see the dotted curve
in Fig.~1.  Alternatively, it might appear
that a judicious increase of the valence quark distributions by
about 10\% in the region $x \simeq 2 E_T/\sqrt{s} \sim
0.35-0.45$
and $Q^2 \sim 10^4-10^5$~GeV$^2$ would remove the discrepancy
for $E_T
> 200$~GeV.  However, these distributions are constrained,
through GLAP evolution, by deep inelastic measurements at lower $Q^2$.
In particular, precise large-$x$ structure function data from BCDMS
($F_2^{\mu p}, F_2^{\mu d}$) \cite{bcdms}, CCFR
($F_2^{\nu N}, x F_3^{\nu N}$) \cite{ccfr} and NMC
($F_2^{\mu n}/F_2^{\mu p}$) \cite{nmc} pin down the dominant $u$ and $d$
distributions very accurately, see for example Ref.~\cite{mrsa},
with combined statistical and systematic
errors of the order of a few percent. Data at smaller $x$ also,
through the effects of evolution and sum rules, constrain the
structure functions at larger $x$.
The situation is further complicated since an increase of the value
$\alpha_S$ (which increases the jet subprocess cross section) tends, through
evolution, to decrease the valence, and hence the jet,
distributions at large $Q^2$.

It is clear that the jet data must be included in a global
analysis to investigate whether a compromise overall fit can be achieved
by adjusting the partons and $\alpha_S (M_Z^2)$, such that we have a
satisfactory QCD description of the detailed shape of the entire
jet spectrum without destroying the fit to the other data.  We
therefore perform a next-to-leading (NLO)
analysis of the available deep-inelastic and related data as
described in Ref.~\cite{mrsa}, but updated to include the published
measurements of $F_2$ at HERA \cite{hera} together with the CDF
jet data \cite{cdf}.  The inclusion of the HERA data impose an
important constraint on the gluon at small $x$ (and rule out,
for example, partons such as MRS(D$_0^\prime$) \cite{mrsd} which were found
to give an excellent prediction of the jet shape for $E_T < 200$~GeV). 
 The HERA data pin down the quarks and gluons at small $x$
and thus (through the sum rules) restrict their variation at
larger $x$.

We fit to the shape of the jet data in the
region $E_T > 50$~GeV where the NLO QCD description is, to a
good approximation, scale independent (and where the experimental
ambiguities from the underlying event are small \cite{cdf}).  To
implement an economical NLO description of the inclusive jet
distribution within the global fit, we first calculate 
the $K$ factor,
\begin{equation}
K \; = \; \frac{\sigma_{LO} \: + \: \sigma_{NLO}}{\sigma_{LO}}
\label{eq:a2}
\end{equation}
as a function of $E_T$ using MRS(A$^\prime$) partons, but with
larger $\alpha_S (M_Z^2) = 0.12$.  The renormalization and
factorization scales are set equal to $E_T/2$.  A global analysis
is performed in which the jet data are fitted to the distribution
$K \sigma_{LO}$.  The resulting optimum set of partons together
with the corresponding value of $\alpha_S (M_Z^2)$ are then used
to
recompute the $K$ factor, and the global fit repeated.  This
iterative procedure converges rapidly. 
For the above choice of scales the $K$ factor is found, to a good
approximation, to be independent of $E_T$ in the interval
$50 < E_T < 400$~GeV and equal to 1.10.
 The whole analysis is
repeated taking the renormalization and factorization scales
equal to $E_T$.  It is found to yield essentially the same global
fit, but the normalization of the jet data increases.

The new set of partons\footnote{A similar fit, but with the jet
data described by $\sigma_{LO}$ was obtained in 
Ref.~\cite{kjs}.}, which we denote by J, is found to have a
significantly larger value of the QCD coupling, $\alpha_S (M_Z^2)
= 0.120$, as compared to 0.113 of MRS(A$^\prime$), in order to
give an improved description of the precise CDF jet data with
$E_T < 200$~GeV, see Fig.~1.  The improvement is obtained at the
expense of a somewhat less satisfactory description of the scaling
violations of the fixed-target deep-inelastic data.  This is
evident from the $\chi^2$ values listed in Table~1.
As an example, Fig.~3 shows the description of the
large $x$ BCDMS $F_2^{\mu p}$ deep inelastic data \cite{bcdms}.
(A similar plot exists for
the $F_2^{\mu d}$ and $F_2^{\nu N},xF_3^{\nu N}$ data.)
Note that there is in fact a continuum of such fits, spanned by A$'$ and J,
in which $\alpha_s(M_Z^2)$ increases from 0.113 to 0.120
for which the fits to the deep-inelastic (low $E_T$ jet data)
gradually deteriorate (improve). In selecting fit J we are
deliberately giving more weight in the fit to the CDF data.

\begin{table}[htbp]
\begin{center}
\begin{tabular}{|l|c|c|c|c|} \hline
      &           & MRS(A$^\prime$) & J     & J$^\prime$  \\
Expt. & \# data & $\chi^2$         & $\chi^2$ & $\chi^2$   \\
\hline
 HERA $F_2^p$ & 206 & 115 & 96 & 95  \\  \hline
 BCDMS & & & &  \\
 $F_2^{\mu p}$ ($x < 0.2$) & 46 & 53 & 43 & 65  \\
 $F_2^{\mu p}$ ($x > 0.2$) & 128 & 204 & 251 & [20703]  \\ \hline
 CCFR & & & &  \\
 $F_2^{\nu N}$ ($x < 0.2$) & 29 & 60  & 74 & 74  \\
 $F_2^{\nu N}$ ($x > 0.2$) & 51 & 35  & 32 & [1290]  \\
 $xF_3^{\nu N}$ ($x < 0.2$) & 29 & 18 & 19 & 18  \\ 
 $xF_3^{\nu N}$ ($x > 0.2$) & 51 & 47 & 40 & [896]  \\ \hline
 NMC & & & &  \\
 $F_2^{\mu p}$ ($x < 0.2$) & 29 & 77 & 74 & 70  \\
 $F_2^{\mu p}$ ($x > 0.2$) & 55 & 36 & 41 & [78]  \\
 $F_2^{\mu p} / F_2^{\mu n}$ & 85 & 137 & 136 & 131 \\ \hline
 CDF & & & & \\
 $d\sigma/dE_T$ ($E_T <$ 200 GeV) & 24 & [158] & 38 & 35 \\
 $d\sigma/dE_T$ ($E_T >$ 200 GeV) & 11 & [14] & 19 & 8 \\ \hline
 \multicolumn{2}{|l|}{$\alpha_S (M_Z^2)$} & 0.113 & 0.120 & 0.129  \\
 \multicolumn{2}{|l|}{CDF (renorm.)} & [1.0] & 1.04 & 1.21  \\ \hline
\end{tabular}
\end{center}
\caption{$\chi^2$ values for various subsets of data; the values
shown in square brackets correspond to data that are omitted from the
fit.}
\end{table}
The compromise fit, J, does not improve the description of the
CDF jet data with $E_T > 200$~GeV.  In fact we see from Fig.~1
that the fit is marginally inferior to A$^\prime$.  It turns out
to be impossible to obtain a reasonable description
simultaneously of the deep inelastic and CDF jet data over the
full range of $E_T$.  Because of their smaller errors the low
$E_T$ CDF data points have a bigger influence on the fit than
those at large $E_T$.  To illustrate the conflicting requirements
of these data we successively omit deep inelastic
data\footnote{However, we keep the NMC measurements \cite{nmc} of
$F_2^n/F_2^p$ in the fit.} starting from those at the largest $x$
values until we can achieve a satisfactory description of the
entire CDF jet $E_T$ spectrum.  We find that it is necessary to
omit data with $x > 0.2$, that is a major fraction of the deep
inelastic data, before a reasonable fit to the jet spectrum is
obtained.  The partons obtained in this way are denoted by
J$^\prime$.  Although the fit to the jet $E_T$ spectrum is
satisfactory (see Fig.~1), these partons give a poor description
of the omitted deep inelastic data --- see, for example, the
J$^\prime$ prediction for the BCDMS data shown in Fig.~3.

In the J$'$ fit the quarks are drastically modified at large $x$
compared to MRS(A$'$)
in order to accommodate the large $E_T$ CDF jet data. The question
addressed in Ref.~\cite{jh} is whether the quarks can be kept essentially
unchanged and the large $E_T$ excess explained solely by a modified
gluon density at large $x$. We have also   performed fits of this type 
and find that the $E_T$ distribution cannot be adequately described 
below and above $E_T \sim 200$ GeV simultaneously -- in qualitative
agreement with the conclusion one draws from the dotted curve \cite{jh} of
Fig.~1. We have also explored the possibiliity of modifying the scale 
for $\alpha_S$ by including a factor $(1-x)/x$, which helps give the jet cross
section a rise with increasing $E_T$. However, the resulting fit can produce
qualitative agreement only over a very limited region ($E_T > 200$~GeV)
provided a large renormalization of the jet data is made.

Fig.~4 compares the quark and gluon distributions, at $Q^2 = 10,
\; 10^4$~GeV$^2$, of the new sets with those of MRS(A$'$).
We plot the singlet quark distribution, $\sum_i (q_i + \bar q_i)$,
since this is most closely related to the combination which is sampled
by the jet cross section. Note that the area under the curves is
the total momentum fraction carried by the quarks and gluons.
We see that the main difference between the A$'$ and J distributions
comes from the different value of $\alpha_S$: the evolution between
$Q^2 = 10$~GeV$^2$ and $10^4$~GeV$^2$ is more pronounced for the J
partons, such that at the latter scale the J quarks are respectively
smaller and larger than the A$'$ quarks for $ x \gapproxeq 0.1$
and $x \lapproxeq 0.1$. This difference leads to a better fit to the
small $E_T$ CDF data, see Fig.~1. The starting gluons are very similar
in the two sets.
For both the A$'$ and J partons the overall renormalization
required to give the best fit to the CDF jet data
is well inside the experimental uncertainty.
The J$'$ partons, on the other hand, have significantly harder quark
distributions at large $x$, as anticipated from Figs.~1 and 3.
At $Q^2 = 10^4$~GeV$^2$, the J$'$ quarks are larger than the A$'$ quarks
for $ x \gapproxeq 0.35$ and $x \lapproxeq 0.08$, giving an improved
fit to the CDF jet data both at large $E_T$ {\it and} at small $E_T$.
Notice also that the J$'$ gluon is significantly smaller than the A$'$
gluon at small $x$ and $Q^2 = 10$~GeV$^2$. This is required in order to
maintain a good fit to the HERA measurements of
$\partial F_2 / \partial \ln Q^2 \sim \alpha_S g$
with a significantly larger $\alpha_S$.

In summary, we have attempted to incorporate the CDF inclusive
jet cross section data in a global parton distribution analysis.
We find that it is impossible to accommodate both the jet data
{\it over the complete $E_T$ range} and the deep inelastic structure
function data.\footnote{This conclusion assumes that the systematic errors
of the CDF data are strongly correlated and that the statistical errors
define the {\it shape} of the distribution. It is relevant to note
that for the UA2 inclusive jet data \cite{ua2jet} there is no evidence
for a discrepancy with the shape of the predicted NLO QCD distribution
(i.e. a ``break" in the region of $x_T \equiv 2E_T/\sqrt{s} \approx 0.25$),
although in this case the large combined statistical and systematic errors 
may mask such structure; see Fig.~5 of Ref.~\cite{jh}.}
However, the CDF data for $E_T < 200$ GeV
and the deep inelastic
data {\it can} be reasonably well fitted simultaneously. The
main effect of including the jet data is to increase the value of
$\alpha_S$ from the standard deep inelastic value. Interestingly,
this slightly improves the fit to the HERA and low-$x$ fixed target
data. The fit to the large-$x$ fixed data is however considerably
worsened, see Table~1. The result of the combined fit is the J set
of partons. It is impossible to simultaneously fit
 the large $E_T > 200$~GeV jet data. If one gives up the fit to the
small $E_T$ data, one can achieve a reasonble large $E_T$ fit by
adjusting the gluon distribution, as in the analysis of
Ref.~\cite{jh} for example. A fit over the whole jet $E_T$ range
(resulting in the J$'$ set of partons)
yields quarks which are completely incompatible with the large-$x$
structure function data, see Fig.~3, a value for $\alpha_S$ which
is several standard deviations larger than the current world average,
and a renormalization of the CDF jet data which is on the edge
of the allowed range \cite{cdf}.

We therefore conclude that it is unlikely that the difference between the
CDF inclusive jet cross section data and the standard NLO QCD prediction
can be attributed to a deficiency in our knowledge of parton
distributions.

\section*{Acknowledgements}
We thank Al Goshaw and Anwar Bhatti for valuable discussions
concerning the CDF data.

\newpage
\section*{Figure Captions}
\begin{itemize}
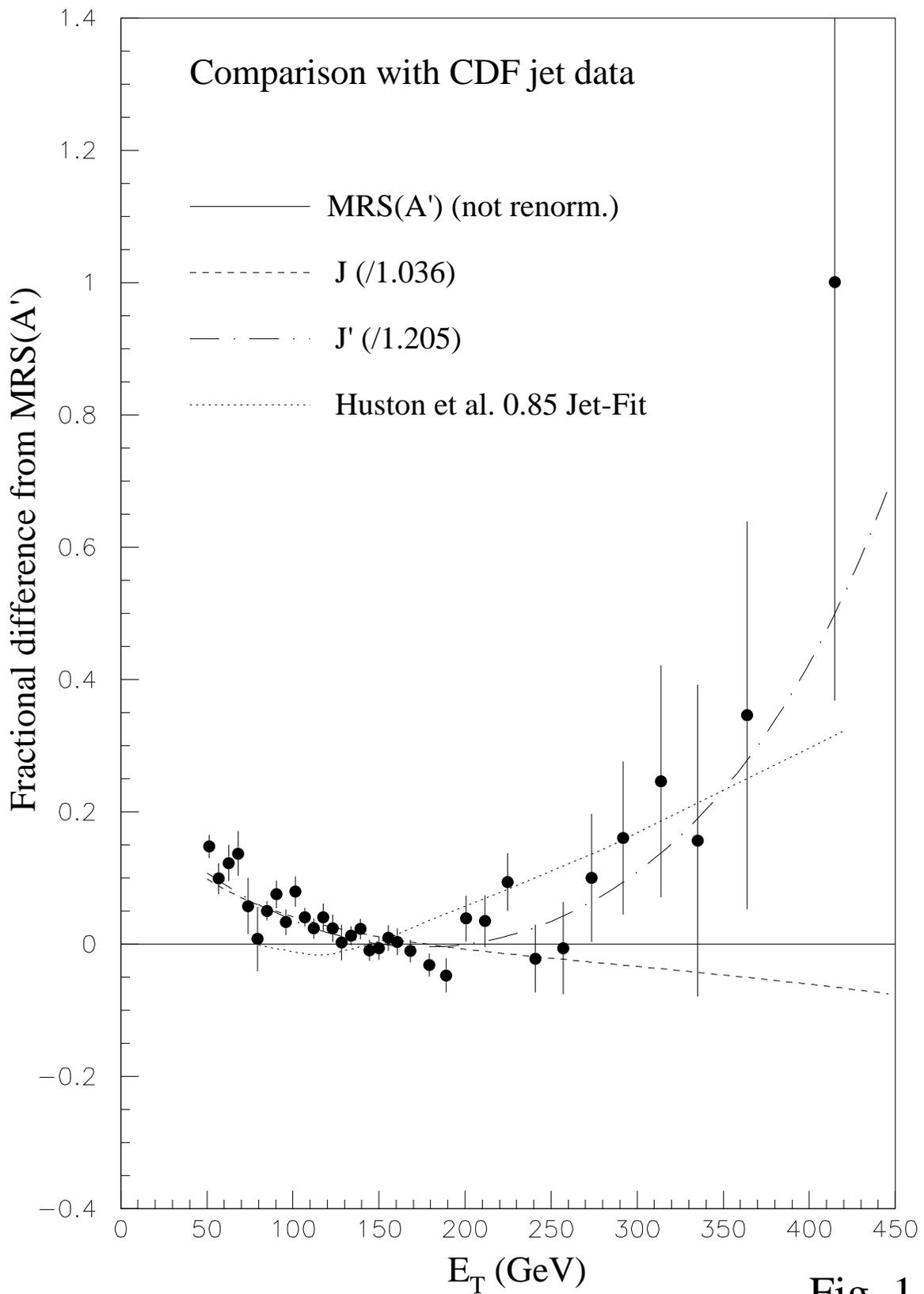
\item[Fig.~1] The fractional difference between the measured CDF
inclusive jet cross section \cite{cdf} and the NLO QCD
predictions based on MRS(A$^\prime$) \cite{mrsag}, J and
J$^\prime$ parton sets.  Also shown (by the dotted curve) is the
gluon-modified description of Ref.~\cite{jh}.
The numbers shown in brackets are the renormalization
factors used for the J and J$'$ partons. Only the statistical errors
of the data are shown.

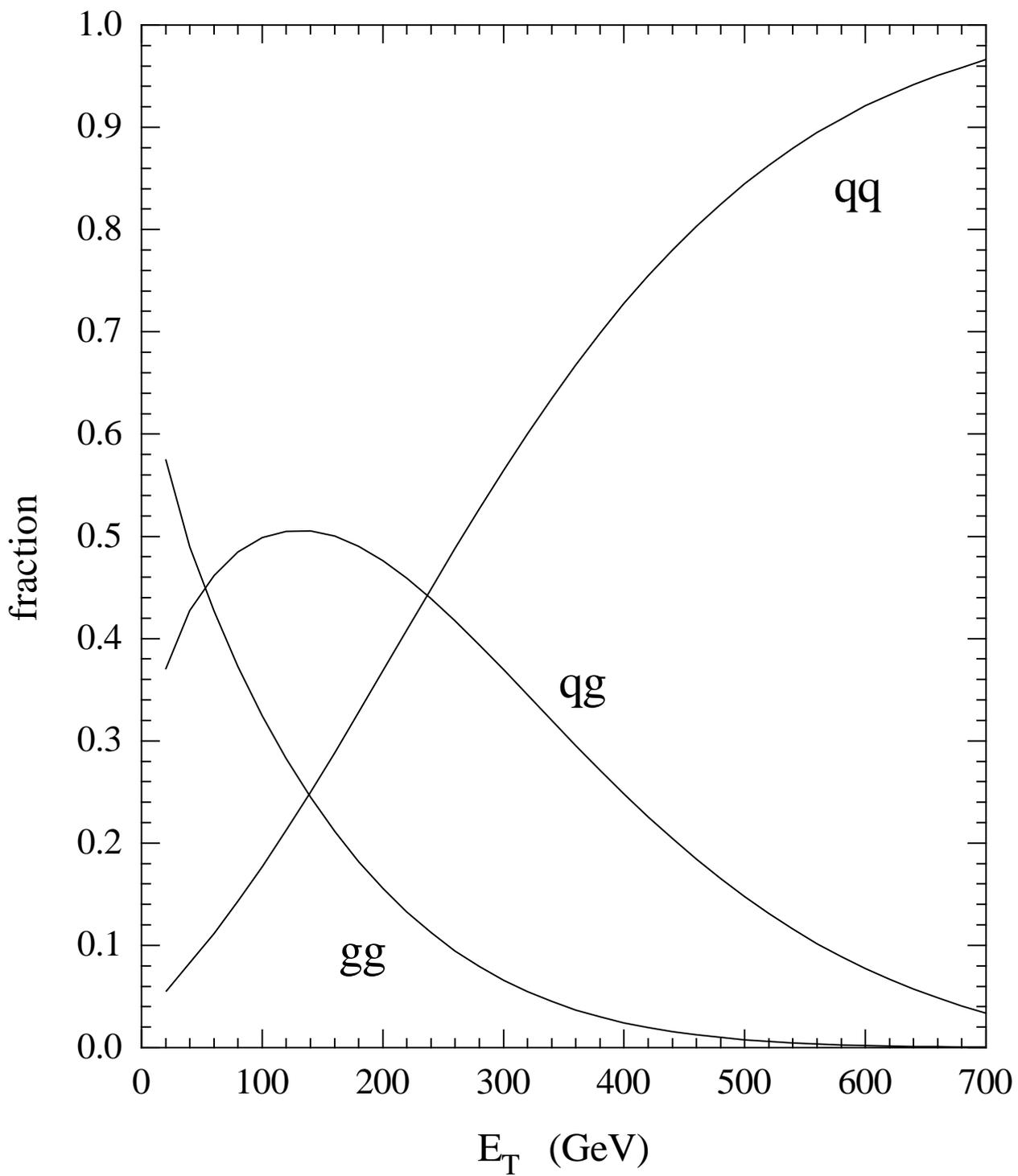
\item[Fig.~2] The fraction of the leading-order
 jet cross section $d^2\sigma/d E_T 
d\eta \vert_{\eta=0}$ at the Tevatron
originating from $gg$, $qg$ and $q\overline{q}$ ($+qq$) initiated QCD
subprocesses using  the MRS(A$^\prime$) set of partons Ref.~\cite{mrsag}.

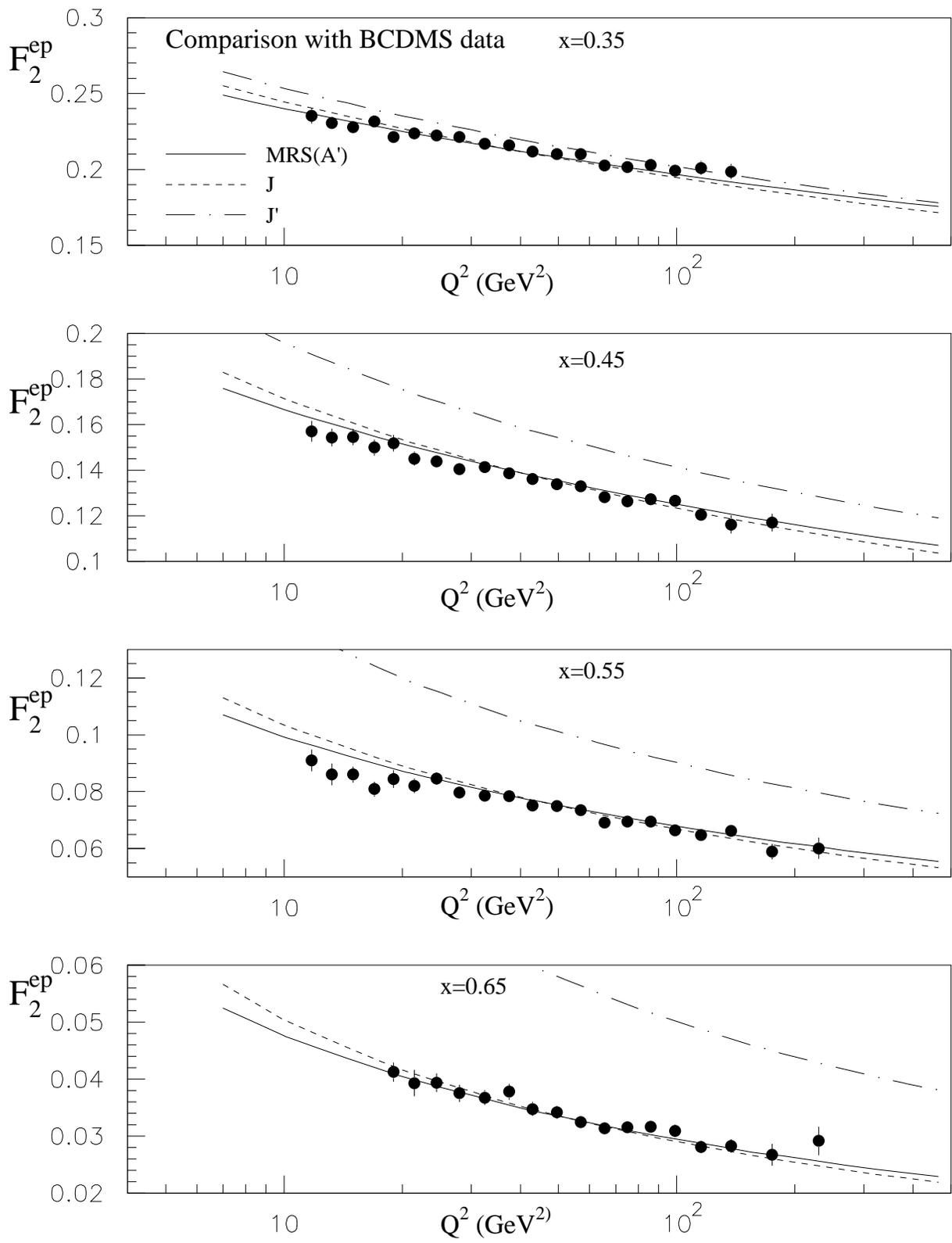
\item[Fig.~3] The description of the large $x$ BCDMS
measurements of $F_2^p$ by the MRS(A$^\prime$) partons of 
 Ref.~\cite{mrsag}, and the J and J$^\prime$ partons of this work.

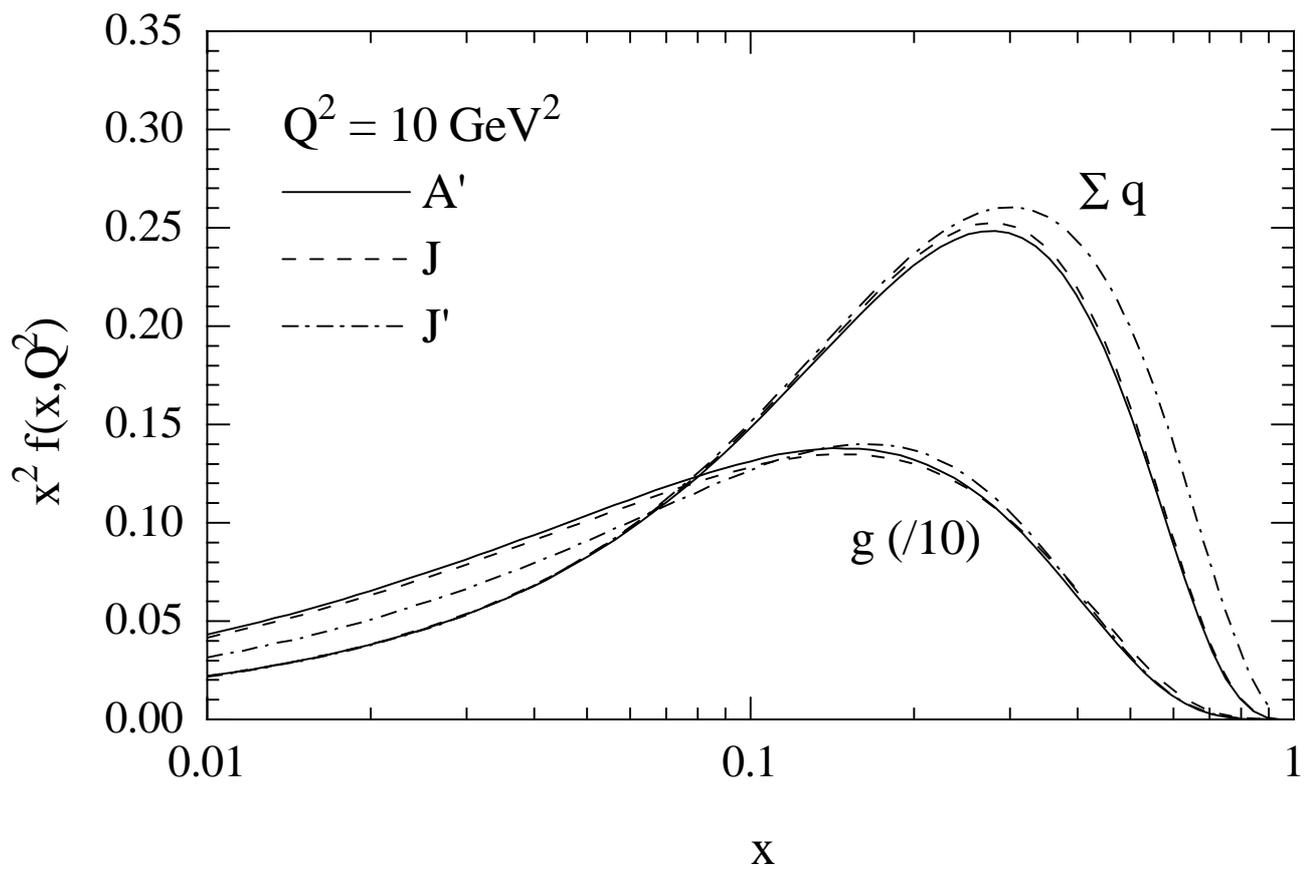
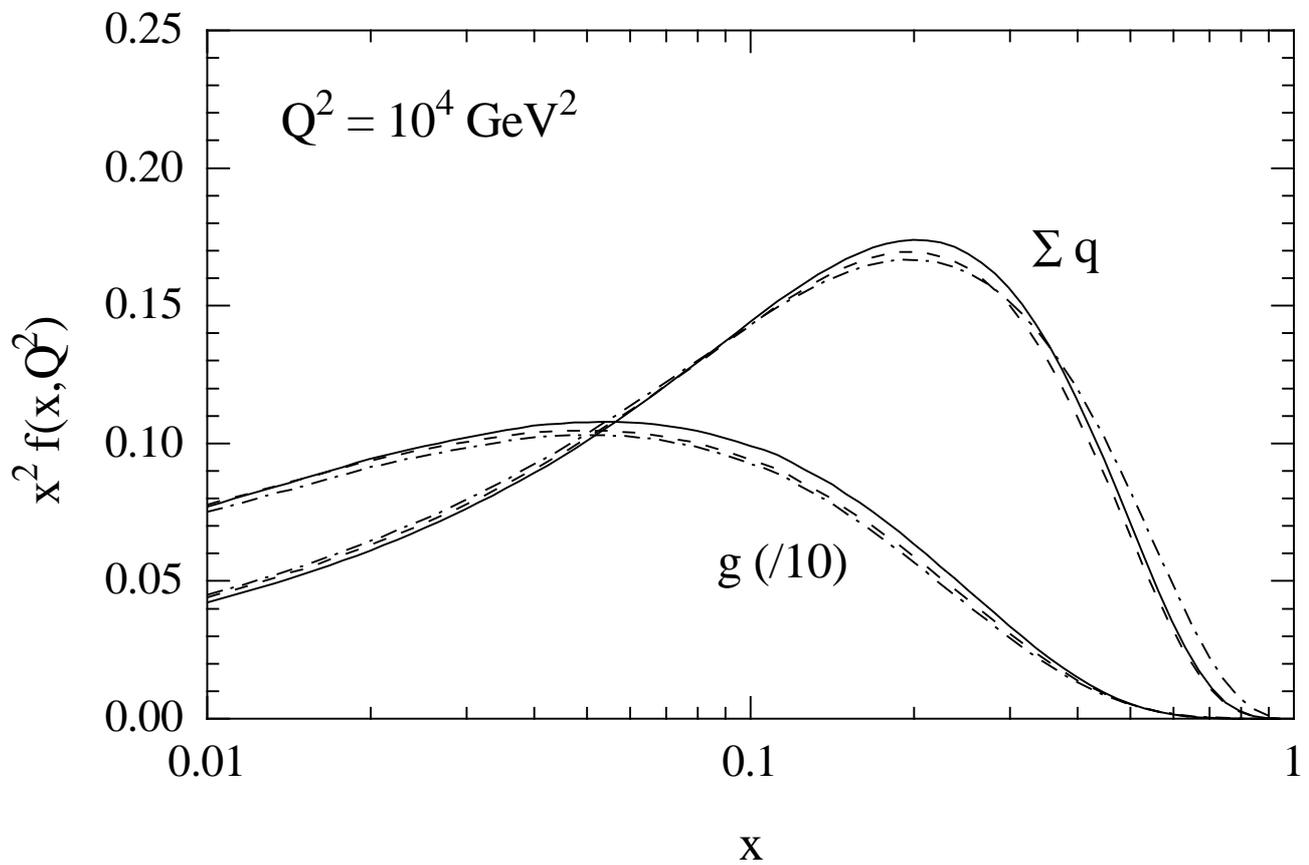
\item[Fig.~4] The parton distributions $g$ and $\sum_i(q_i+
\bar{q}_i)$ at $Q^2 = 10$ and $10^4$~GeV$^2$.
\end{itemize}
\end{document}